# CHALLENGE OF ACHIEVING EFFICIENT SIMULATIONS THROUGH MODEL ABSTRACTION

Hessam S. Sarjoughian

Arizona Center for Integrative Modeling and Simulation
School of Computing, Informatics, and Decision Systems Engineering
Arizona State University
699 S. Mill Avenue, Tempe, AZ 85281, USA

William A. Boyd

Arizona Center for Integrative Modeling and Simulation
School of Computing, Informatics, and Decision Systems Engineering
Arizona State University
699 S. Mill Avenue, Tempe, AZ 85281, USA

Miguel F. Acevedo

Electrical Engineering Department and
Advanced Environmental Research Institute (AERI)
1155 Union Circle #310440
Denton, TX 76203, USA

**ABSTRACT**

Coupled natural systems are generally modeled at multiple abstraction levels. Both structural scale and behavioral complexity of these models are determinants in the kinds of questions that can be posed and answered. As scale and complexity of models increase, simulation efficiency must increase to resolve tradeoffs between model resolution and simulation time. From this vantage point, we will show some problems and solutions by using as example a vegetation-landscape model where individual plants belonging to different species are represented as collectives that undergo growth and decline cycles spanning hundreds of years. Collective plant entities are assigned to cells of a static, two-dimensional grid. This coarse-grain model, guided by homomorphic modeling ideas, is derived from a fine-grain model representing plants as individual objects. These models are developed using Python and GRASS tools. A set of experiments is devised to reveal some barriers in modeling and simulating this class of systems.

## 1 INTRODUCTION

Understanding the dynamics of some natural systems is only practical in simulated settings. An example is studying coupled human-environment interactions during the Neolithic period (Barton, et al., 2016). Developing this type of model poses a number of challenges including the amount of details that can be realistically modeled so that we can simulate the models over many hundreds to thousands of years. The implication is that although, for example, plants representing vegetation in a landscape have different characteristics and variability at an individual level, they need to be modeled at high levels of abstraction, without all of their biological processes represented (Acevedo, et al., 1999). Modeling leaves, branches, trunks, and roots for many individuals of many plant species while examining different landscape and climate processes is impractical. Similarly, modeling ecological processes for individual plants at high temporal resolution (say daily) is far too computationally demanding when very large number of plants occupy extensive land surfaces and we are interested in vegetation changes spanning many hundreds of years.

Questions that are of interest to ecological anthropologists require simulation over thousands of hectares which may then include millions of plants if treated individually. Therefore, while it can be useful to develop high resolution vegetation models to understand ecosystem dynamics, it may be unnecessary having fine-grain details for plants where small changes may have limited or no measurable impact on human decision making. This is, in part, due to challenges that anthropologists, ecologists, and geologists encounter in accessing or obtaining complete and/or accurate data sets from the distant past.

In this paper, we will show a coarse-grain Vegetation-Landscape (VL) model that is derived from a fine-grain model representing individual plants of several plant species or functional types. The simplified or coarse-grain model can be simulated a few to several orders of magnitude faster while retaining some, but not all, characteristics of interest that are observed in the fine-grain model. We will develop one way of assigning individual plants to collectives aided by homomorphic modeling principles. The specific assignment, although important, it is not the main objective of this paper. To further reduce execution time, we have a basic algorithm that parallelizes simulation, using multiple processing cores available in desktop computing platforms. The resulting VL model is evaluated relative to its fine-grain counterpart. Our goal is to show that models that share common characteristics, as the VL fine-grain and coarse-grain models, do not easily lend themselves to homomorphic modeling.

## 2  BACKGROUND

When developing coupled natural-human (CNH) model, a purpose for including models of vegetation dynamics over a large landscape with heterogeneous terrain is to help understand interactions between human decisions and natural landscape processes. Barriers facing simulation of vegetation includes not only the dynamic complexity of individual plants, but also the areal extent of the landscape. Instead of modeling changes in land topography, a land surface representing 2D spatial grid representation can be used. In this paper, we focus on vegetation-landscape modeling and not with modeling land erosion and deposition. High resolution models of a few hundred individual plants over a few hundred grid-cells are practical to simulate for several hundred cycles in a reasonable amount of time. Efficient implementation of this fine-grain model can take a few minutes on a basic desktop computer. At a scale of millions of plants, simulation can take many hours, or even weeks on a high-end, multicore desktop computer (see Section 5). This is because this kind of simulation requires storing and processing very large amounts of data. The entire state of the simulation is tracked and updated to produce results. As is the case in this kind of study, the processing and memory requirements associated with high resolution simulations can make such simulations impractical unless the model can be seamlessly ported to be simulated on thousands of processors.

### 2.1  Fine-grain Vegetation-Landscape model

The fine-grain VL is akin an agent-based model (Acevedo M., 2015) to serve as the vegetation-landscape model for a CNH model to understand land-use change during the Neolithic period in the Mediterranean, specifically the southern Spain landscape (Barton, et al., 2016). This fine-grain VL is based on equations and algorithms of standard gap models but applied over a landscape. Simulation of this model allows tracking changes to plants within a stand and ported to the landscape by assuming that the plants respond to terrain characteristics such as elevation, soil, and geomorphic properties, and summarized as abstracted terrain types such as ridge, slope, and valley. A stand is represented as a cell (which vegetated or not) of a grid representing the landscape. The simulation progresses through a set number of time steps, each representing one year. At each time step, the model uses a land surface or terrain map generated as output by the geomorphic dynamics model component of the CNH model, and maps representing fire and other human actions on the landscape generated as output by the agent based model representing the human system component of the CNH model. The species of Mediterranean vegetation are abstracted as four functional groups following (Pausas J. G., 1999) as resprouter and fire-intolerant (type 1), resprouter and fire-tolerant (type 2), non-resprouter and fire-tolerant (type 3), and non-resprouter and fire-intolerant (type 4). The fine-grain model (Acevedo M., 2015) considers so far types 1 and 3, which are factored with two lifeforms (tree and shrub) to further abstract four vegetation types exemplified by emblematic species of each type: Type 1, Tree - Resprouter and fire-intolerant. Example: *Quercus,* Type 1 Shrub - Resprouter and fire-intolerant. Example: *Erica,* Type 3, Tree - Seeder and fire-tolerant. Example: *Pinus*, Type 3, Shrub - Seeder and fire-tolerant. Examples: *Cistus* (Acevedo M., 2015). During each time step, for each of these Quercus, Erica, Pinus, and Cistus vegetation types, the number of germinating and dying plants along with growth are determined for each landscape cell. The outputs are a number of maps for each time step, with each map representing the values of variables (by species and stand summary) of the vegetation such as basal area, density, biomass, leaf area index, seed bank, dead biomass. These VL model output maps are passed to the human component of the CNH for human agent decision making (e.g., burn, clear land) and to the geomorphic component of the CNH to determine land erosion at each time-step. Figure 1 shows the data relationships and calculations for individual plants (trees) by vegetation type ('species') in each cell,

Currently, the fine-grain model determines the dynamics of individual plants of each vegetation type ('species') within this cell independently of dynamics at other cells.

The parameter values for terrain and plant species are defined by files read before initialization. When the simulation begins, each grid-cell in the landscape is initialized with a random number of plants of various basal diameter and age for each species. For each time step, updates are made to the states of the plants at each grid-cell (see Figure 2). First, the model accounts for the plants killed by fire for each species, and the growth rates for the surviving plants are calculated. Then, the model determines which plants die from natural causes. The attributes of the surviving plants are updated based on growth rate. The number of germinating plants for each species is found and used to update the number of plants and seed bank. After all of these changes are made, output maps are produced to show the state of the vegetation at the end of the time step. Some of the calculations, particularly for determining the growth rates of the plants, depend on the states of the other plants in the cell, as well. The number of the terrain grid-cells and the number of plants that must be updated determine time and complexity for the VL simulation for some given computing software/hardware platform (see Figure 1). The VL is implemented in Python and can interact with GRASS for creating, manipulating, and displaying maps.

The VL model is implemented to process and produce GRASS GIS raster maps. GRASS does not perform any computations within the VL simulation. A raster map corresponds to the landscape grid. Each grid-cell of a raster map can hold a value. Each raster represents a specific attribute that varies with location, such as the amount of sunlight at the location or the number of vegetation types.

While detailed individual-based simulation results of this model are useful to analyze natural-human interactions using CNH for small landscapes (say hundreds of ha), the simulation time is too long for large landscapes (say thousands of ha) using high-end personal computers. It is helpful to note that while execution efficiency of the fine-grain VL model improves through use of multiprocessing, the gained efficiency is not satisfactory for large landscapes. Although, the fine-grain model already includes abstraction of species and terrain, it is useful to seek further abstraction of the individual-based dynamics to improve its simulation efficiency for large landscapes.

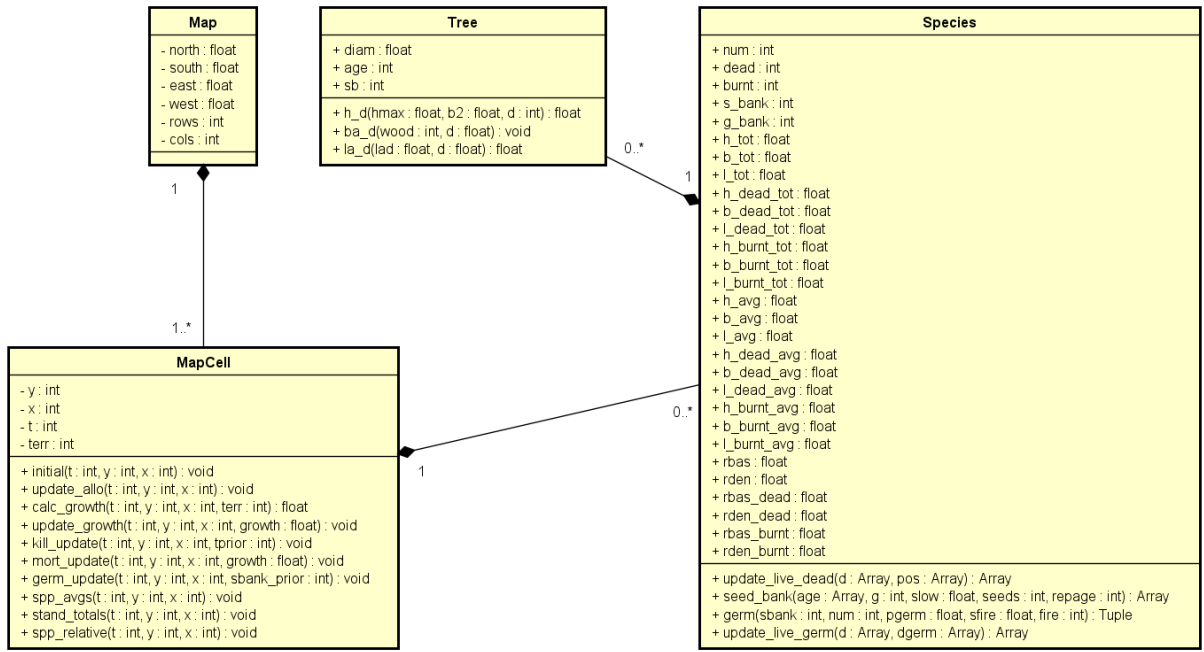

Figure 1: Data relations for the fine-grain VL model

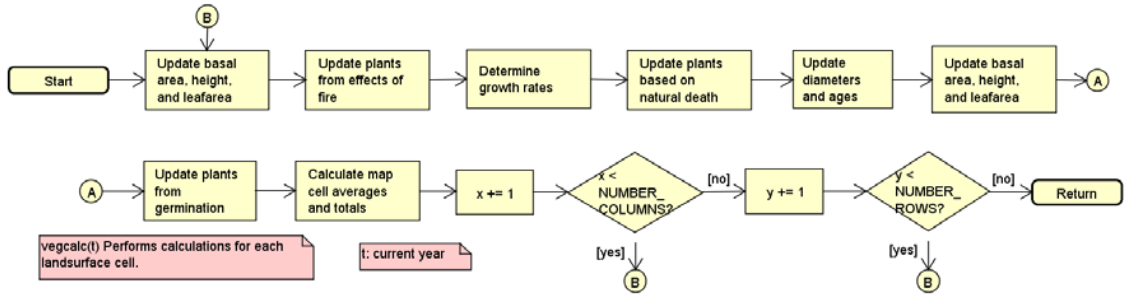

Figure 2: A common procedure for computing changes in the state for any of the four plant types

## 2.2 Homomorphic modeling

Ideally, the relationships between data sets and functions for two different model abstractions can be achieved using homomorphic modeling as illustrated in Figure 3 (Wymore, 1993; Zeigler, Praehofer, & Kim, 2000). For the class of discrete-time models, for a given high-resolution and low-resolution models, it might be possible to find input, state, and output homomorphic mappings. These mappings show exactly how input, state, and output sets from one model maps to those of another. To guarantee functional homomorphism (state and output changes for input changes), state and output homomorphic consistency relationship also have to be found (Davis & Bigelow, 1998). Obtaining these homomorphic mappings and consistency relationships, however, can be challenging even when models are developed using discrete-time, continuous-time, or discrete-event modeling formalisms. With careful data and function abstractions, it may be possible to derive lower-resolution (coarse-grain) dynamics from high-resolution (fine-grain) models with acceptable tradeoffs between simulation accuracy and execution time. Establishing homomorphism between models requires simulating both models and therefore exhaustively showing that their state and output trajectories are consistent according the state transition and output functions consistency relationships. However, in some cases it is impractical to simulate fine-grain models even for a limited set of experiments. This is the case for the fine-grain VL model described in Section 2.1.

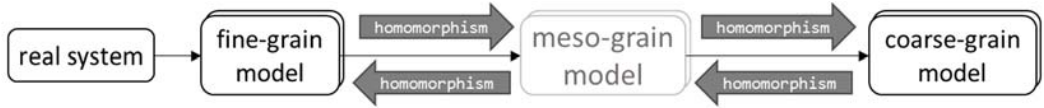

Figure 3: Model abstractions with homomorphic and consistency mappings

## 3    RELATED WORK

The concepts of model simplification and elaboration has been well studied in abstract settings (e.g., (Zeigler, Praehofer, & Kim, 2000)). It has been applied to various domains with varying degrees of success (e.g., (Davis & Bigelow, 1998)). An abstraction for multiresolution is defined in terms of time, object, process, and space. A model is at a higher resolution, for example, if its data and functions are more has more details relative to another model, one that has less data and simpler functions.

For vegetation-landscape modeling several approaches to develop coarse-grain landscape from fine-grain models have been implemented. Formal procedures to combine forest species into functional roles and to derive a semi-Markov landscape model from a gap model based on species or on those species roles have been developed (Acevedo, Urban, & Ablan, 1995) (Acevedo, Urban, & Shugart, 1996; Abbot-Wood, 2002). In addition to determining the next state from based on transition probabilities, as in Markov models, semi-Markov models also use the amount of time in each state to determine the next state. These procedures are formalized as algorithms to determine probabilities from fine-grain simulations and have been implemented for various forest landscapes using a variety of computational platforms including clusters (e.g.,(Acevedo, et al., 1999)).

The semi-Markov model becomes then a 'metamodel' that computes efficiently for large landscapes. Alternative methods to derive and parameterize metamodels from fine-grain gap model include cellular automata and stage-structured projection matrix approaches (Urban, Acevedo, & Garman, 1999). Similar projection matrix metamodels have been developed for grassland vegetation (Raventos, Segarra, & Acevedo, 2004). For Mediterranean vegetation-landscapes, three approaches (Pausas & Ramos, 2006). Two of these approaches define landscape coarse-grain dynamic rules applied to functional types as conceptual analog to those rules at small areas or fine-grain. The third approach is to implement a gap model over the landscape, in the same fashion to the VL fine-grain model described in section 2.1.

## 4    COARSE-GRAIN VEGETATIONLANDSCAPE  MODEL

Given the fine-grain VL model, a next level of abstraction, is to consider a canonical plant that is a representative of a collection of individuals of the same vegetation type belonging to an individual grid-cells (Figure 4). In a naïve way, this canonical plant is an "average" plant or a representative of a cohort of identical individuals, as employed early on to abstract gap models (Swartzman & Kaluzny, 1987) are implemented in R-packages for forest simulation (Acevedo M., 2012). This method is also similar to applying functional type rules at the landscape cells (Pausas J., 2006) which also include cell interaction. Comparing Figures 1 and 4, it can be observed that "Tree" is not modeled in the coarse-grain VL model. This affects calculations for growth rate and leaf areas, which are performed for a collective entity, not individual plants. Such a collection of plants is assigned to each grid-cell for each vegetation type. To avoid introducing more complexity the coarse-grain vegetation-landscape model, retains the assumption of independent process at each grid cell as in the fine-grain VL model. That is, the dynamics of the plants in each grid-cell are not affected by the plants assigned to all other neighbor grid-cells.

## 4.1 Vegetation data variables

As described in section 2.1, the fine-grain model tracks diameter, age, number of seeds, height, and basal area for each plant of each map cell of the model. Multidimensional arrays are used to store these data. Within each cell, separate array indices are assigned for each species and for each plant of a species (see Figure 4). For three of the variables (diameter, age, and number of seeds), the values must be preserved between time steps, so the arrays are stored on disk for every row and column of the map. Other attributes are also determined at every map cell, but are not preserved after the computations for a cell are completed for a time step.

Since some attributes are stored for each of the plants of the map, the data complexity for a fine-grain simulation is no less than $\Theta(total\ number\ of\ trees) = \Theta(r, c, s, m)$, where $r$ is the number of map rows, $c$ is the number of map columns, $s$ is the number of vegetation types, and $m$ is the maximum number of plants for one vegetation type in one map cell. This data-complexity imposes a lower limit for the time-complexity of the simulation, because the attributes are updated for each of the plants at each time step. The data-complexity for the coarse-grain VL model depends on aggregate attributes for species of each map cell. For example, the average age of a species of plants in a cell are represented. Since attributes for each individual plant are not needed, the data-complexity is independent of the number of plants and thus results in increasingly more efficient simulations as the number of plants in a landscape is increased.

## 4.2 Vegetation functions

In the fine-grain VL model, the state of all plants assigned to a landscape grid-cell define the state of the grid-cell's vegetation. As a discrete-time model, in each time step, therefore, the state of every plant is updated.. The fine-grain VL, in contrast, does not track the state of every plant. Only the aggregate values for each of the species assigned to a grid-cell are read and set at each time step. In other words, the procedure shown in Figure 2 cannot account for collections of plants instead of individual plants. This is because the calculations from the fine-grain model cannot be simply repeated given the difference in the two model abstractions shown in Figures 1 and 4.

The coarse-grain VL model uses these functions such that they satisfy state change and output function consistency relationships relative to the fine-grain VL model, in conjunction with some well-defined input, state, and output homomorphic mappings. This approach to coarsening a fine-grain model allows state changes in the coarse-grain model to be consistent with the state changes of the fine-grain model. With an output function consistency relationship, the fine-grain and coarse-grain models are homomorphic to one another. Alas, these models are not homomorphic to one another as defined. Instead, some functions of the fine-grain models are reformulated in view of changing the state of a collection of plants, rather than a single plant.

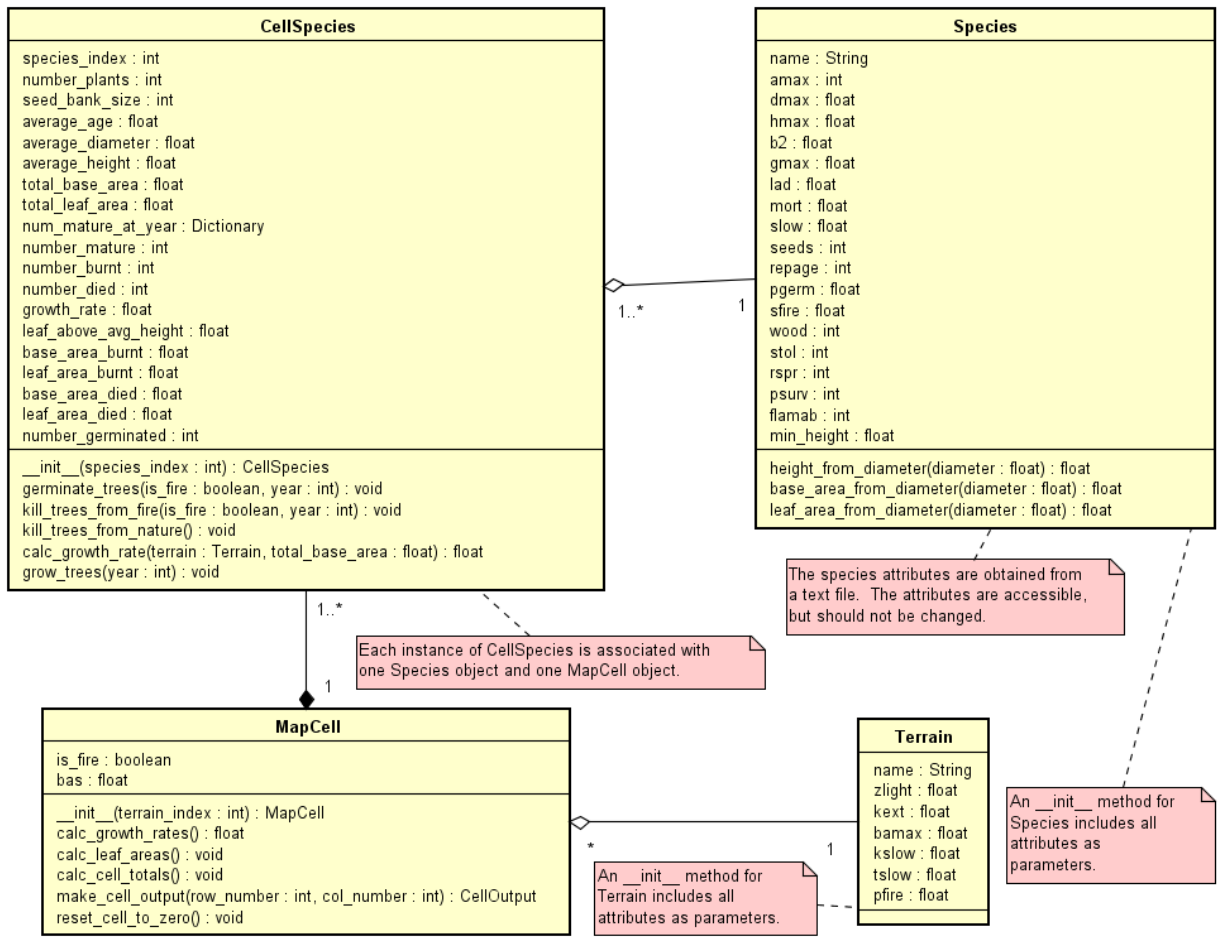

Figure 4: Coarse-grain VL model

As shown in Table 1, calculations for attributes such as plant diameter, height, and leaf area are significantly different. As shown in Figure 5, plant diameter for five plants can be averaged. Summing and averaging diameters for the plants and increasing by a constraint value to the average value is identical to increasing the plants' diameters individually and then averaging them. This is a simple, linear mapping. In the case of death of plants (e.g., due to natural causes or occurrence of fire), the mapping is nonlinear and imperfect as shown in Figure 5. The average age of surviving plants is an example of an amount that could not be determined using operations similar to those used in the fine-grain version. In the fine-grain model, each plant has its own age. The plants that die are determined probabilistically, and older plants are more likely to die of natural causes. Although this does not change the ages of the surviving plants, the average age of surviving plants is likely to change. In the coarse-grain vegetation model, the average age after the deaths of plants is determined probabilistically by finding the average age of the plants that died, using a normal distribution. The mean of the distribution is set slightly higher than the current average age, to model the greater likelihood of older plants dying.

In general, for functions that perform nonlinear changes to individual states, there are no counterpart functions that can achieve the exact effect desired when applied to aggregate states. For example, in the fine-grain model, the survival of a plant is determined probabilistically, while in the coarse-grain model the number of collective survivors is determined probabilistically. These observations show that formulating a coarse-grain model from the fine-grain vegetation model using homomorphism as described in Section 2.2 is not possible.

Table 1: Examples of mappings for calculating plant attributes

| Calculations | Fine-grain model | Coarse-grain model |
|---|---|---|
| Height | $h = h_{max} * (1 - e^{(-hd_a * d)})$ | $h = h_{max} * (1 - e^{(-hd_a * d_{ave})})$ |
| Basal area | $Area_{base} = \left(\frac{\pi}{4}\right) * d^2$ | $Area_{base_{total}} = \left(\left(\frac{\pi}{4}\right) * d^2_{average}\right) * plants_{number}$ |
| Leaf area | $leaf_{area} = leaf_{area_{dead}} * diamater^2$ | $leaf_{area_{total}} = (leaf_{area_{dead}} * diameter^2_{ave}) * number_{trees}$ |
| Leaf area above a given height $h$ | Sum of areas of leaves of every plant taller than $h$ from each species | $\left(leaf_{area} * \left(1 - \frac{h_{min}}{h_{ave}}\right) / (h_{min} - h_{max})^2\right) * (h_{min} - h_{max})^2$ |
| Growth rate | $g_{max} * spaceFactor * lightFactor * respFactor$ | $g_{max} * spaceFactor * lightFactor * respFactor$ |
| Diameter update | $d = d + growth$ | $d_{total} = d_{total} + growth * number_{plants}$ |
| Age update | $age = age + 1$ | $age_{ave} = age_{ave} + 1$ |
| Tree fire survival | Age does not change | $age_{ave} \leftarrow (age_{ave} * plants_{original_{number}} - age_{dead} * plants_{dead_{number}}) / plants_{next_{number}}$ |
| Seeds in seed bank | If $age \geq age_{adult}$ then $S_{bank} \leftarrow C_{seeds}$ else $S_{bank} \leftarrow 0$ | $S_{bank} \leftarrow plants_{mature_{number}} * C_{seeds}$ |

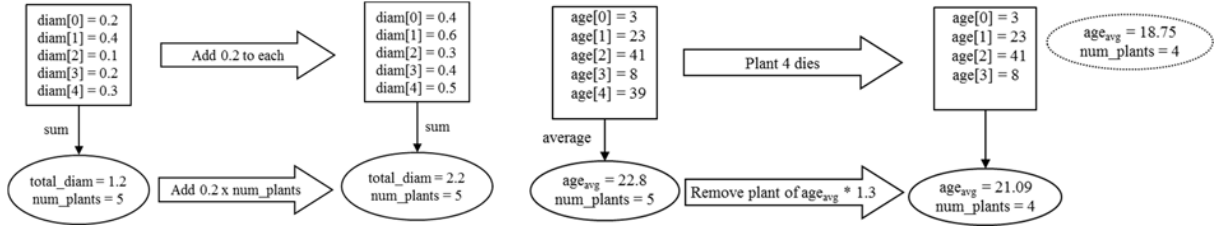

Figure 5: Perfect and approximate exemplar mappings

## 5 EXPERIMENTS

The two fine-grain and coarse-grain VL models are simulated within the same simulator. The same input values – terrain map and fire locations – are provided to both versions of the simulator, in the same formats. The simulator has separate data sets and functions. Since there is no data sharing between the versions, both models are simulated concurrently and independently. Separate outputs are produced from identical inputs for each of the experiment configuration shown in Table 2.

Table 2: Experiment configurations

| Total # of cells | Max. plants per species per cell | Ave. initial # of plants per cell | Random fires per year | # of years | # of null cells | Cell size | Map size |
|---|---|---|---|---|---|---|---|
| 4 | 100 | 20 | No | 200 | 3 | 100 m2 | 20m x 20m |
| 5,000 | 100 | 20 | Yes | 200 | 0 | 100m2 | 1000m x 500m |
| 20,000 | 100 | 20 | Yes | 200 | 0 | 25m2 | 1000m x 500m |

**Single cell with maximum of 100 plants per vegetation type**

| Terrain | Density | Basal area |
|---|---|---|
| Ridge | | |
| Slope | | |
| Valley | | |

Figure 6: Plant density and basal area for each terrain type

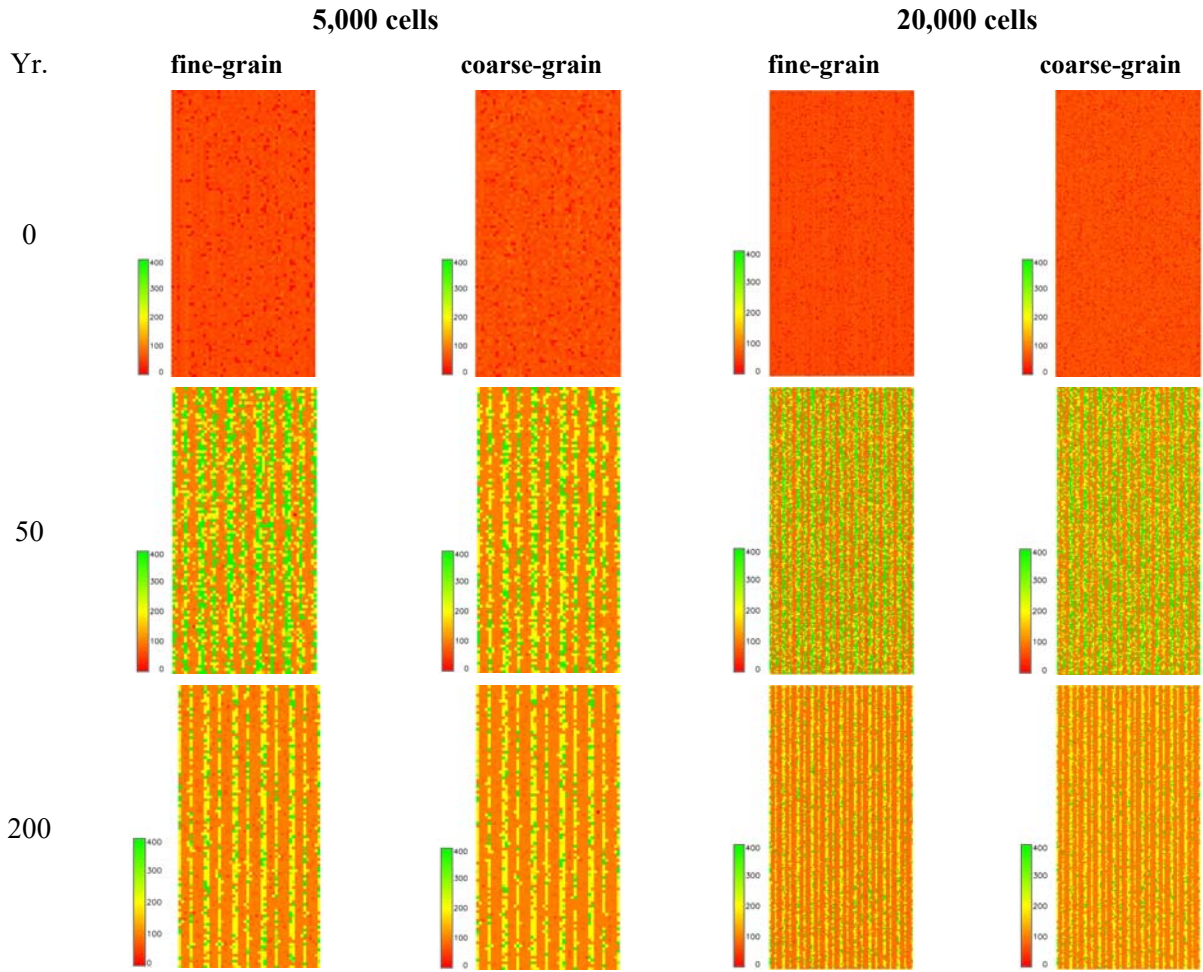

Figure 7: Plant density maps at different years

In order to closely observe the results of the change in abstraction on each grid-cell, VL was configured to run using a single active grid-cell. The map also included three null-value grid-cells, which do not have a terrain type and do not hold any plants. Separate single grid-cell simulations were run using each of three terrain types: ridge, slope, and valley. The total plant density (total number of plants in a grid-cell) and the total basal area (in square meters, added for all plants of a cell) were observed over 200 simulated years. The maximum number of plants allowed for each species was varied in different simulations. Four different vegetation types were defined for use in all the simulation trials. Graphs for the simulations allowing 100 plants per species in a grid cell are shown in Figure 6.

Plant density shows very similar patterns from the fine-grain and coarse-grain versions of the simulation; the plant density increases rapidly until the maximum of 100 plants per species (400 total) is reached. The plant density then remains at the maximum value, indicating that the germination rate remains at least as high as the death rate for each time step. The results for basal area are greatly different in the fine-grain and course-grain simulations. The basal area in the course-grain version tended toward much greater values. This suggests that the method of computing the basal area from the average diameter is not a reliable approximation for the process of finding the sum of individual basal areas.

Simulations were run using the same configuration for maps with 5,000 cells and 20,000 grid-cells of various terrain types, as shown in Figure 7. In these simulations, however, fires were set to occur in each grid-cell, with a probability depending on the type of terrain. The plant densities showed very similar behavior to fire in the fine-grain and coarse-grain simulations, with the difference between the total plant densities produced averaging to less than 1%. As in the single-cell simulations, the basal area values from the coarse-grain simulations did not approximate those of the fine-grain simulation, sometimes reaching values more than 5 times as high as the fine-grain basal areas.

Because the coarse-grain adaptation of the simulator worked with a much smaller data set, it was able to complete the simulations using much less time. For the simulations with 5,000 grid-cells, the running time for the coarse-grain version was about 5.9% of the time required for the fine-grain simulation. The increase in performance is a result of the reduced number of state updates that needed to be performed for the smaller, coarse-grain data set.

However, the large discrepancy in results for basal area indicates that for these non-linear systems we need to search for an alternate way of consistently aggregating from fine-grain to coarse-grain. Future work can explore use of cohorts of identical individuals (Swartzman & Kaluzny, 1987; Acevedo M., 2012), probabilistic and matrix metamodels (Acevedo, Urban, & Shugart, 1996; Urban, Acevedo, & Garman, 1999), and functional types rules (Pausas & Ramos, 2006). Another line of research is to approach this challenge through use of some suitable modeling formalism where homomorphic input, state, and output mapping relationships and state change and output functions (Wymore, 1993; Zeigler, Praehofer, & Kim, 2000) can be formalized instead of relying on similarities between results.

All the simulations were run using a Dell Precision T3610 computer with an Intel Xeon 3.70GHz 8-core processor. PyCharm Community Edition 5.0.1 was used to edit and run the software, which was interpreted using Python 2.7.6 and NumPy 1.8.2. Output maps were converted for display using GRASS 7.0.3.

## 6  CONCLUSION

In simulation-based study of multifaceted dynamics, such as those of CNH systems, it may be necessary to tradeoff model abstraction with computation time and resource needs. In this paper, we exemplified some problems and solutions when attempting to reduce the structure and behavior of fine-grain models while still answering some questions of interest. We showed one way of abstracting out some details of a VL model. In particular, a coarse-grain model was derived from a fine-grain VL model. Our goal was to explore the concept of abstracting one model from another and note that applying formal homomorphic modeling may not be feasible. We devised a variety of experiments, the results of which shed light on pragmatics of multi-resolution modeling. This research, we hope, can help in furthering research in multi-resolution modeling. This, in turn, we anticipate to lead toward building simulations that can help studying larger and more complex systems.

**ACKNOWLEDGMENTS**

This research is supported by NSF grant #DEB-1313727. We would like to thank C.M. Barton and I.I. Ullah for fruitful discussions.

**REFERENCES**

Abbot-Wood, C. (2002). *Landscape Forest Modeling of the Luquillo Experimental Forest, Puerto Rico.* Masters Thesis, University of North Texas, Denton, TX. Retrieved from digital.library.unt.edu/ark:/67531/metadc3362/
Acevedo, M. (2012). *Simulation of ecological and environmental models.* Boca Raton, Florida: CRC Press, Taylor & Francis Group.
Acevedo, M. (2015). *Mediterranean vegetation model.* http://medland.asu.edu.
Acevedo, M. F., Pamarti, S., Ablan, M., Urban, D. L., & Mikler, A. (2001). Modeling Forest Landscapes: Parameter Estimation From Gap Models Over Heterogeneous Terrain. *Simulation, 77*(1-2), 53-68.

Acevedo, M. F., Urban, D. L., & Ablan, M. (1995). Transition and Gap Models of Forest Dynamics. *Ecological Applications, 5*(4), 1040-1055.
Acevedo, M. F., Urban, D. L., & Shugart, H. H. (1996). Models of forest dynamics based on roles of tree species. *Ecological Modelling*(1-3), 267-284.
Barton, C. M., Ullah, I. I., Bergin, S. M., Sarjoughian, H. S., Mayer, G. R., Bernabeu-Auban, J. E., . . . Arrowsmith, J. R. (2016). Experimental Socioecology: Integrative Science for Anthropocene Landscape Dynamics. *Anthropocene, 13*, 34-45.
Davis, P. K., & Bigelow, J. H. (1998). *Experiments in Multiresolution Modeling (MRM)*. RAND/MR-1004-DARPA, Rand Corporation.
Pausas, J. (2006). Simulating Mediterranean landscape pattern and vegetation dynamics under different fire regimes. *Plant Ecology, 187*, 249–259.
Pausas, J. G. (1999). Mediterranean Vegetation Dynamics: Modelling Problems and Functional Types. *Plant Ecology, 140*(1), 27-39.
Pausas, J. G., & Ramos, J. I. (2006). Landscape analysis and simulation shell (LASS). *Environmental Modelling and Software, 21*, 629-639.
Raventos, J., Segarra, J., & Acevedo, M. F. (2004). Growth dynamics of tropical savanna grass species using projection matrices. *Ecological Modelling, 174*(1-1), 85-101.
Swartzman, G., & Kaluzny, S. (1987). *Ecological Simulation Primer.* MacMillan.
Urban, D. L., Acevedo, M. F., & Garman, S. L. (1999). Scaling Fine-scale Processes to Large-scale Patterns using Models derived from Models: Meta-Models. In D. J. Mladenoff, & W. L. Baker, *Spatial modeling of forest landscape change: Approaches and applications* (pp. 70-98). Cambridge, UK: Cambridge University Press.
Wymore, A. W. (1993). *Model-based Systems Engineering* (Vol. 3). Boca Raton, FL: CRC press.
Zeigler, B. P., Praehofer, H., & Kim, T. G. (2000). *Theory of Modeling and Simulation: Integrating Discrete Event and Continuous Complex Dynamic Systems.* Academic press.

## AUTHOR BIOGRAPHIES

**HESSAM S. SARJOUGHIAN** is an Associate Professor of Computer Science and Computer Engineering in the School of Computing, Informatics, and Decision Systems Engineering (CIDSE) at Arizona State University (ASU), Tempe, AZ, and co-director of the Arizona Center for Integrative Modeling & Simulation (ACIMS). His research interests include model theory, poly-formalism modeling, collaborative modeling, simulation for complexity science, and M&S frameworks/tools. He is the director of the ASU Online Masters of Engineering in Modeling & Simulation program.
He can be contacted at sarjoughian@asu.edu.

**WILLIAM A. BOYD** is a Computer Science Masters student at ASU, Tempe, AZ, USA. He can be contacted at waboyd @asu.edu.

**MIGUEL F. ACEVEDO** is a Regents Professor in the Electrical Engineering Department and a member of the Advanced Environmental Research Institute (AERI) at the University of North Texas, Denton, TX, USA. He can be contacted at Miguel.Acevedo@unt.edu.